# PERFORMANCE EVALUATION OF STRUCTURED AND SEMI-STRUCTURED BIOINFORMATICS TOOLS: A COMPARATIVE STUDY

Raja A. Moftah[1], Abdelsalam M. Maatuk[2], Richard White[3]

[1,2]Faculty of Information Technology, Benghazi University, Libya

[3]School of Informatics, Cardiff University, UK

## ABSTRACT

*There is a wide range of available biological databases developed by bioinformatics experts, employing different methods to extract biological data. In this paper, we investigate and evaluate the performance of some of these methods in terms of their ability to efficiently access bioinformatics databases using web-based interfaces. These methods retrieve bioinformatics information using structured and semi-structured data tools, which are able to retrieve data from remote database servers. This study distinguishes each of these approaches and contrasts these tools. We used Sequence Retrieval System (SRS) and Entrez search tools for structured data, while Perl and BioPerl search programs were used for semi-structured data to retrieve complex queries including a combination of text and numeric information. The study concludes that the use of semi-structured data tools for accessing bioinformatics databases is a viable alternative to the structured tools, though each method is shown to have certain inherent advantages and disadvantages.*

## KEYWORDS

*Bioinformatics, Sequence Retrieval System, Entrez, Perl, BioPerl.*

## 1. INTRODUCTION

A large number of biological databases developed by bioinformatics' experts, which contain extensive information related to nucleic acid are made available over the World Wide Web (WWW). The ultimate purpose of the bioinformatics field is to facilitate the discovery of biological developments by means of the creation and maintenance of databases for specific biological information [1]. However, the rapid increase of this type of databases implies not only database design issues, but the increase of complex interfaces, where the users or researchers can access existing data [1]. Meanwhile, scientists working in various fields of bioinformatics use available data for research purpose. To access these data from the internet, search engines can be utilised as data retrieval tools [3]. Several data retrieval tools require the ability to perform searches to obtain bioinformatics data required by scientists in their studies and analysis. Within the field of bioinformatics, databases generally attempt to arrange biological data into meaningful relationships, while simultaneously developing approaches to retrieve useful information from these relationships that employ effective structures for both search and analysis of the biological data. The design of high quality databases can be achieved through the use of a variety of formats that range from various types of data to diversity approaches, where researchers may want to use this data [4]. Thus, biological data represented within bioinformatics are stored according to the internal structures of specific databases.

Data retrieval tools are easily available to efficiently access information from the internet. These tools allow the retrieval of information through various easily used search criteria [15]. To benefit from data stored in bioinformatics databases, the users requirements should be met, in order to enable easy extraction of data required to answer specific biological question [5]. At present, a great deal of bioinformatics research is concerned with the technology of both hardware and software used to create, manage and use databases. These technologies can include such features as common repositories of gene data, such as GenBank database [6, 19], along with other private databases, such as those used by research groups involved in gene mapping projects, or by biotechnology companies. Currently, the majority of researchers use many public databases, where data related to such biological structures such as nucleic acids or protein sequences, along with their associated complex information types, such as networks and pathways that are deposited, manipulated and stored [7].

This paper presents an evaluation of methods used to access bioinformatics databases through approaches that include accessing public structured and semi-structured data. We have used Sequence Retrieval System (SRS) and Entrez as tools to retrieve information as structured data, and Perl and BioPerl as tools to retrieve information as semi-structured data from public bioinformatics databases. This study focuses on two bioinformatics databases, i.e., European Bioinformatics Institute (EBI) [2] and National Centre for Biotechnology Information (NCBI) [8]. These databases provide improved data access to users through their services, which are available all the time, and are associated with conventional databases, such as relational transactional systems. The main reason for using structured databases is the fact that they facilitate data discovery and provide full structured information in response to queries. We have applied these tools in retrieval of biological data, and then give the corresponding results, supported by samples of the gene accession number. The results of these tests are presented, along with a discussion of how these results fit within the study context, exploring the differences among them along with the recognized advantages and disadvantages of using these tools.

## 2. BACKGROUND AND RELATED WORK

Several methods and tools exist to extract data from biological databases. These methods retrieve bioinformatics data stored in structured and semi-structured database systems. This section throws light and investigates these databases and methods.

### 2.1 BIOINFORMATICS DATABASES

The most important collections of bioinformatics databases that are currently available without restriction on the internet include both EBI and NCBI databases. The services of these databases meet the needs of their users by enabling them to access data and associated software tools freely, along with allowing data to be available for download. Both databases achieve the support of their users through either the management of their requirements, and allowing them to find assistance by visiting the Web. For example, the EBI database performs access to various sources by EB-eye and SRS, whereas the NCBI database allows access through Entrez and the eUtils tools.

Biological raw data are accrued in public bioinformatics databanks, such as Genbank [6] and EMBL [2], where they are subsequently classified according to their respective properties before being stored in specialized sub-databases [9]. Databases are designed to assist users to query them more efficiently while gaining more accurate accessing to information. This data, which are often held within several different databanks or databases, are frequently accessed simultaneously while being concurrently correlated with each other using specifically designed user interfaces [9]. However, the way in which this data is queried depends upon the specific algorithm used. For example, the examination of data, or the expert use of analysis tools, can aid researchers in understanding the way, in which these tools are used to analyse data and to infer results. Therefore, it can be seen that data integration can allow researchers to associate or integrate relevant data from other databases [4]. Therefore, information has to be readily and easily accessible for researchers to start on their data analysis.

The commonly available methods for data retrieval are browsing and keyword searching, although they have some limitations. Browsing is unsuitable for locating specific items of data, because it relies on the user following links, which can be both tedious and confusing, as it is easy for a user to get lost using this method [10]. Thus, users need tools, which allow them to retrieve information quickly and precisely, retrieving exactly what they need for analyses about the core biological entities, through provision of knowledge about these entities. In a restricted sense, this requires a fast and efficient search engine that provides easy to use and uniform access to the biological data resources hosted on public free bioinformatics databases [11].

## 2.2 SRS AND ENTREZ STRUCTURED DATA RETRIEVAL TOOLS

The data retrieval tools for bioinformatics databases, which are evaluated in this study include SRS and Entrez systems. Both of these tools were designed for purpose of querying with keywords, and used to efficiently gather information from public free access resources available on the internet [12]. These tools can be used to access data that is usually associated with relational databases, in which information is formatted into rows and columns within tables. Most database management systems (DBMS) are constructed for structural data, where data is arranged in entities, where similar entities collected together, as either relations in relational databases or classes in object-oriented databases [17, 29]. In these databases, entities in the same collection have the same attributes, which are same for all entities in a group: they have similar defined configuration, may have a predefined length, and are in the same order in each record.

There is argument over whether SRS supports data structure through providing exact indices for implementing list of sub-entities [19, 16, 17, 20]. It works well with simple keywords, using the extended web form. However, some work argue that the shortcoming of SRS cannot deal with a satisfied way within complex enquires involving numerical data or calculations, and that users need to take long time to deal with the mixture of information [19, 17]. Entrez has also been proven to be a powerful tool with simple queries, but one that has severe limitations with respect to retrieving subs-entities that need to be manually extracted. Entrez allows easy retrieval of information from internal structured data [18, 20]. However, the limitation of this server is poorly adapted because it inefficiently deals with subsequence [17].

## 2.3 PERL AND BIOPERL SEMI-STRUCTURED DATA SEARCH TOOLS

The other type of tools that are covered in this study depend on computer languages, which use bioinformatics databases to deal with biological data, i.e., programming language and scripting. This type of language is generally derived from open sources projects, such as Perl and BioPerl, both of which are widely used in bioinformatics for various applications that include the analysis of DNA sequences and protein sequences. This type of languages deal with biological information and retrieve semi-structured data from internal structured data of a particular resource. The data in this type of structured retrieval is arranged by grouping similar entities together, but entities in the same group might not necessarily have the same attributes, as the order of attributes is not important, not all attributes may be required, and the size or type of attributes in a group may differ [16].

Perl and BioPerl tools used to extract information as text and to compute the numerical process are proved that they are flexible enough to support the retrieval of information in semi-structured form. However, both are able to manipulate and extract information as full structures, which were exploited for performing the aims of this study. Perl gives access to data stores in GenBank database via flexible series of sequences. The high throughput technologies usually require the extraction of large data sets of the data related with sequence information [17]. SRS and Entrez are keywords tools for retrieving entries from GenBank or EMBL databases, which find certain limitations in querying complex keywords. Croce et al. [21] exploited Perl script to deal with same data and improve their work by using the BioPerl to retrieve the same information, and found that Perl and BioPerl scripts are usually necessary to retrieve the exact data search that is easily usable [17].

Stajich et al. [25] state that Perl is extremely successful for connecting software applications together to sequence analysis and extract information structured as text [25]. Perl software is widely used in bioinformatics in laboratory or institution, and that the script through this software was written for immediate utility rather than reusability [25]. This results in considerable inefficiency, for the same software is written various times, leading the scientific community to enhance the Perl script into a BioPerl toolkit to allow reusable Perl modules including generalized routines specific to life-science information. BioPerl is also useful to retrieve information as semi-structured data, because it is able to handle queries without knowing data types and provides flexible querying for data that allows a user to request condition in their queries to enable expected to fetch answer which match with original query. Perl is used for bioinformatics applications because of its speed [27]. In Perl, you can often toss such a program off in few minutes and the research can proceed. This rapid prototyping ability is often a key consideration when choosing Perl. It is common to find programmers familiar with both Perl and C who claim that Perl is five to ten times faster to program in than C. Perl script is regularly necessary in order to retrieve the exact data searched for embed in a user friendly interface [21]. Moreover, Perl scripts can easily be modified, if necessary, by users for unforeseen needs. However, BioPerl is a toolkit of routines to make it easier to develop Perl script and reduce codes in such programs.

Loshin [13] describes structured data as information that can be simply modelled, structured, formed and formatted in ways that are simple for a user to manipulate, whereas a user is familiar with the unstructured text in documents such as articles. Loshin [13] adds that the intermediate nature of semi-structured data, having contained structure, but not enough of a regular structure to "qualify" for the types of management and automation often applied to structured data. Users encounter semi-structured data on a daily basis, both in technical and non-technical settings. For instance, web pages follow certain typical forms, with content implanted within HTML pages usually having other types of metadata within the tags. In semi-structured HTML, the tags that give the semi-structure serve a number of different purposes in terms of either formatting instructions to a browser or in giving reference links to both internal anchors and external pages. This automatically involves exact details about the data to be accessed. A non-technical example would be traffic signs posted along highways, while different areas utilize their own local protocols; it is generally possible to deduce the correct exit after seeing a few signs. The fact that sufficient regularity exists to extract some information makes semi-structured data interesting and potentially very useful [13].

## 3. METHODOLOGY AND IMPLEMENTATION

### 3.1 DESIGN

A realistic example of the type of data retrieval and analysis performed by bioinformatics researchers has been chosen as a basis for demonstrating the effectiveness of structured and semi-structured methods. The first step to deal with the implementation of information retrieval related to gene sequence was to obtain the accession numbers, a sample of which is taken from [14]. The sample provided has been exploited in this study to access appropriate remote databases for the purpose of retrieving a combination of textual and numerical data related to the Authors, Title and Journal for each accession number of nucleic acid, as well as to retrieve the mean and range of gene sequence length through the use of two variables. This investigation only counted the letter Cysteine over 20% in the proteins sequences and also calculated the mean, maximum and minimum of the Cysteine [3, 28].

We have used EBI and NCBI structured bioinformatics databases, by exploiting their search tools, i.e., SRS and Entrez. The retrieved data files are illustrated by accessing data tables on a server and processing the data. The data retrieval is displayed in tables as web pages with a fixed structure and requires users to click on links to fully explore the retrieved data. We used an example of retrieving semi-structured data through the use of tools, such as simple Perl program and BioPerl Toolkit, which is a collection of Perl modules that are the most commonly used programming languages in bioinformatics. The aim of using Perl and Bioperl is to focus users on their specific data requirements rather than reporting all of information associated with their inquiry. In this case, data is presented as a semi-structured data XML text, produced directly by using these programs [18].

## 3.2 RETRIEVING INFORMATION USING STRUCTURED DATA TOOLS

### 3.2.1 RETRIEVING INFORMATION USING SRS TOOL

The first method to be investigated is through the use of a web page, which interfaces with a search program on a database server that holds structured databases. For example, through SRS, a user can link and mine databases within minutes, while EBI stored datasets in SRS that can be interrogated by external programs to perform virtually any computation. The users, here, need to find the EBI website and then click on the relevant databases, to select SRS from the database browsing clean, click on SRS and paste a list of sequence IDs into this. This list should be of the format DATABASE:ID, and can be created by using the Microsoft Excel program to modify the IDs. Further, it is important to ensure that each entry is on a single line and that the database exists on the server. The result reveals after clicking on Search are shown in Fig.1.

| EMBL | Primary Accession (Links to SVA) | Accession List | Description | Sequence Length |
|---|---|---|---|---|
| EMBL:AB003356 | AB003356 | AB003356 | Eel mRNA for estrogen receptor, complete cds. | 3061 |
| EMBL:AB009669 | AB009669 | AB009669 | Anguilla japonica mRNA for eKir, complete cds. | 1119 |
| EMBL:AB012869 | AB012869 | AB012869 | Anguilla japonica mRNA for natriuretic peptide receptor-A, complete cds. | 3896 |
| EMBL:AB015638 | AB015638 | AB015638 | Anguilla japonica CYP1A mRNA for cytochrome P450 1A, complete cds. | 3453 |
| EMBL:AB016041 | AB016041 | AB016041 | Anguilla japonica mRNA for eSRS3, complete cds. | 1385 |
| EMBL:AB016169 | AB016169 | AB016169 | Anguilla japonica mRNA for gonadotropin I beta subunit, complete cds. | 1066 |
| EMBL:AB019371 | AB019371 | AB019371 | Anguilla japonica mRNA for ventricular natriuretic peptide, complete cds. | 1024 |
| EMBL:AB019372 | AB019372 | AB019372 | Anguilla japonica mRNA for atrial natriuretic peptide, complete cds. | 670 |
| EMBL:AB020414 | AB020414 | AB020414 | Anguilla japonica CYP1A9 mRNA for cytochrome P450 1A9, complete cds. | 2786 |
| EMBL:AB023960 | AB023960 | AB023960 | Anguilla japonica mRNA for androgen receptor alpha, complete cds. | 5168 |
| EMBL:AB025356 | AB025356 | AB025356 | Anguilla japonica mRNA for activin B, complete cds. | 3346 |
| EMBL:AB025357 | AB025357 | AB025357 | Anguilla japonica mRNA for proliferating cell nuclear antigen, complete cds. | 1571 |
| EMBL:AB025361 | AB025361 | AB025361 | Anguilla japonica AR-beta mRNA for androgen receptor-beta, complete cds. | 3720 |
| EMBL:AB026989 | AB026989 | AB026989 | Anguilla japonica mRNA for prepro-mGnRH, complete cds. | 379 |

Fig. 1 : Query results using SRS

First, the users need to retrieve the results from the ID query, after which they select each ID to display related information in order to extract information on the Authors, Title and Journal for each accession number of nucleic acid. Retrieving the mean and range of the gene sequence length was also done manually, meaning that a user needs to spend a considerable amount of time to access the final results as the process is done manually.

### 3.2.2 RETRIEVING INFORMATION USING ENTREZ TOOL

Entrze is a powerful tool with simple queries, however, the users need to click to each access number to access a detailed display of information associated with each access number. This means that for each access number, a user needs to find the NCBI [8] website, and then write each sequence access number in the search box to retrieve the related information seeking for. The users then need to type this accession number into another search box, selecting the option to display all the information presented by all databases associated with NCBI. After this, Entrze provides all information that it has been able to find, which is related with this access number various databases, but the users will need to look to extract information associated with gene. Therefore, the users need to click on each nucleic acid record to present the information needed. In addition, the users need to retrieve all information for each access number and gather this separately, then they will be able to manually calculate the mean and range of the gene sequence length as shown in Fig. 2. However, the users can directly retrieve information related to each access number by choosing from the beginning nucleic on the search options, then writing the access number and finally by selecting the search option to display all desired information related with each access number.

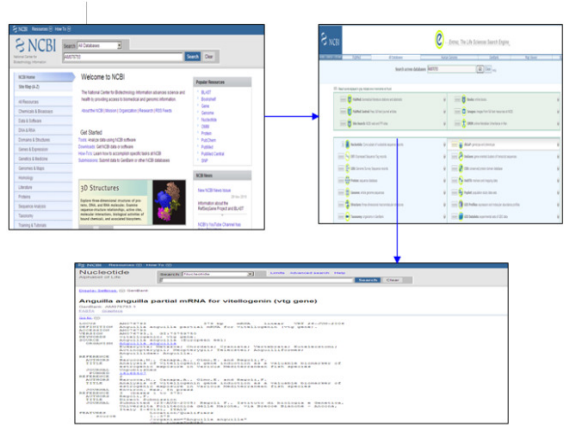

Fig. 2: Query results using Entrez

Retrieving the mean and range of the gene sequence length is done manually, meaning that user needs to spend a considerable amount of time to access the final results as the process is done manually.

### 3.3 RETRIEVING INFORMATION USING SEMI-STRUCTURED DATA TOOLS

#### 3.3.1 RETRIEVING INFORMATION USING PERL TOOL

In order to access data, e-utility tools were exploited, a piece of software first posted as an E-utility URL to NCBI. After this has been retrieved, subsequent to which it processes the data as required. The software can therefore use computer language to send a URL to E-utilities server

32

and interpret the XML response, using Perl code to demonstrate its use in semi-structured data retrieval. NCBI Entrez system is used again, but in this case the eFetch Utility is used to retrieve data as a semi-structured mixture of information text and numeric statistics. The purpose of this program is to access appropriate remote databases, i.e., GenBank, using the same accession numbers to search on a set of genes and their proteins. The program was modified to use eFetch to fetch information needed. The query result for this code is shown in Fig. 3.

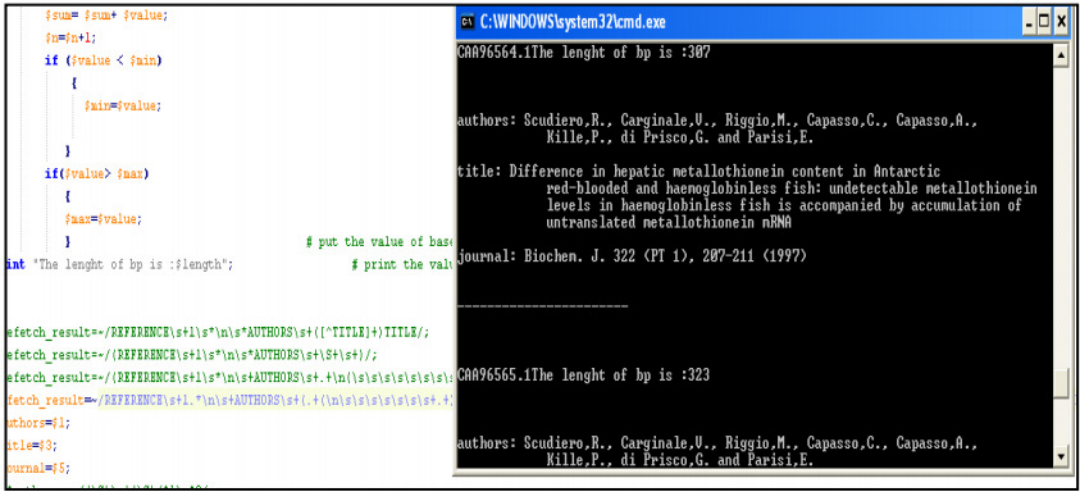

Fig.3 : The results from using Perl program

The program represents the mean and range of the nucleic sequence length by using two variables, setting the minimum to a larger value than any possible value, and the maximum to a value smaller than any possible value. This means that when the main program loop is first executed the program compares the value with the current value of minimum and maximum along with use of exchange operating and measures the mean. According to the nucleic which include 285 sequences:

- The mean of the lengths for the genes is: 1093.466666667
- The maximum is: 5423
- The minimum is: 77.

All these information were obtained from GenBank database by providing the IDs of the genes of interest. GenBank was used to extract only the sequences of the proteins, counting the number of times. The letter C occurs in the amino acid sequence. We use an "if" conditional statement to decide if the value is more than 20%, returning the results shown in Fig. 4. In addition, the program shows the mean, maximum length, and minimum length of the nucleic acid.

International Journal of Software Engineering & Applications (IJSEA), Vol.9, No.5, September 2018

Fig. 4: Shows the results from using Perl program

This program also presents the mean and the maximum length and minimum length of proteins which contain only Cysteine and display these values as:

- The mean length of the Cysteine-rich sequences is: 375.679012345679
- The maximum length is: 1756
- The minimum length is: 105.

There are many differences between the whole set of genes and the genes that have coded proteins including high proportions of Cysteine in their amino-acid sequences, that were Metallothioneins. Differences could also be seen in the mean, maximum, and minimum sequence lengths. The proteins, which includes large proportions of Cysteine had a smaller mean length than the whole set of 285 sequences of genes, and the maximum for example for 285 genes in total is 5423, which is more than the maximum for the protein Metallothionein, which includes the Cysteine at a value of 1756.

Perl has been shown to use powerful tools to retrieve specific data rather than the full information related to the user inquiry, when compared with SRS and Entrze. Further, Perl was able to deal with specific data required by the users and can be programmed quickly to achieve this, as well as allowing data to be easily saved to a personal computer.

### 3.3.2 RETRIEVING INFORMATION USING BIOPERL TOOL

BioPerl is not a new language but instead is a collection of Perl modules that facilitate the development of Perl scripts for bioinformatics applications. We have used BioPerl script to retrieve the same information already retrieved using SRS, Entrez and Perl script. Perl program is converted using BioPerl script to achieve the same results. In addition, we have provided command line for all code. This study has used BioPerl modules, which include: Bio::DB::Eutilities, which is a module imported to fetch the details for the provided list of protein IDs and used to populate the genbank.db file; and Bio::SeqIO, which is a module that enables the extraction of information from structured data in the genbank.db. Further details can be found in [15, 20] in order to understand the BioPerl script, its modules and commands for retrieving details from databases. The pseudo code of this program is given in Fig 5.

*Read the list of Gene IDs from list.txt and store in an array*
*Using Bio::DB::EUtilities, fetch the details from protein db and store in benbank.gb file*
*Read the file genbank.gb using Bio::SeqIO*
*Check for the sequence length and store information in variables for sum, minimum length and maximum length*
*Fetch and display the values of variables Authors, Title and Journal*
*Check for the number of Amino Acids and calculate percentage then check the value of percentage is more than 20% then compute sum, minimum length and maximum lengths for such ids*
*Display the results for Average Length, min and max length and store the same in report output.txt*
*End.*

Fig 5: The pseudo code of data retrieval program

By using Bio::DB::EUtilities, the users can fetch the details of the gene IDs in the file. The following command produces the new object $factory, which refers to the protein database for the IDs stored in array @ids.

*my $factory = Bio::DB::EUtilities->new(-eutil => 'efetch',*
*-db    => 'protein',*
*-rettype => 'gb',*
*-id    => \@ids);*

The next commands start retrieving details from the database and store them in a file. As a value of $file is genbank.gb, after this section, the user will have all details in the file and can read it to gather the information required:

*$factory->get_Response(-file => $file);*

By using SeqIO module, the users can read the file and get the required information:

*my $seqin = Bio::SeqIO->new(-file => $file, -format => 'genbank'*

The users can also get the sequence individually from the file using this command:

*my $seq = $seqin->next_seq*

After the users get the sequence, they can use Perl's basic functions to find the required details. Length function gives the length of the sequence string. The users will then need to fetch the details for all sequences to calculate the minimum, maximum lengths. Variables will be used to store the minimum and maximum values and their final value will be the final result.

*if ($lengthSeq< $minLen) {*
*$minLen=$lengthSeq;*
*}*

*if($lengthSeq> $maxLen){*
*$maxLen=$lengthSeq;*
*}*

Using the annotations on the reference, the users can get the values of title, author and journals in this section as Perl script result as shown in Fig. 6.

*$annotation->get_Annotations("reference")*
*$ref->authors(),$ref->title(),$ref->location()*

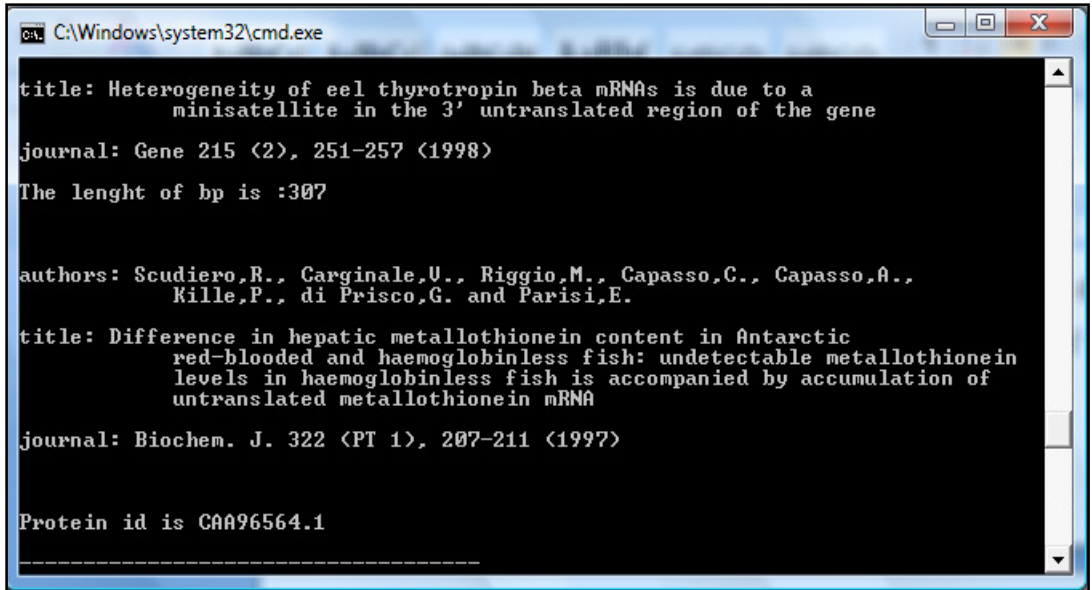

Fig. 6: The results from using BioPerl program

Using $seq->get_SeqFeatures the users can get the other features of the sequence response, the values of which are stored in ref $feat_object in the program.

*$feat_object->get_tag_values('protein_id')*
*$feat_object->get_tag_values('translation')*

The first command gives protein ID and the next one gives the protein sequence and can be used to find the number of amino acids. The following command shows how the authors counting number of 'C's in the sequence

*$cc =($translation=~ tr/C//);*

Also this program calculates the percentage of amino acid and stores this in the variable $gas, as well as checking whether it is more than 20 percent or not. The maximum and minimum length of the sequence are stored where the percentage is more than 20%.

*if ($valueCC< $min20)*
*{*
*$min20=$valueCC;*
*}*
*If ($valueCC> $max20)*
*{*
*$max20=$valueCC;*
*}*

The final step is to print the values and display them on the screen, as shown in Fig. 7.

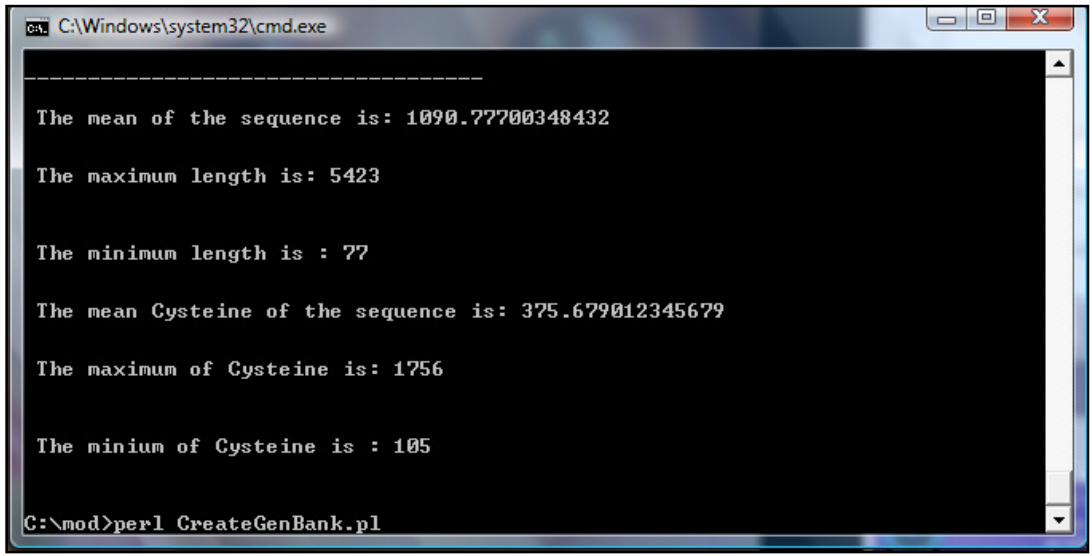

Fig. 7:  The results from using Perl program

### 3.4 TOOLS EVALUATION

This section describes the evaluation and compassion of SRS, Entrez, Perl and BioPerl. Firstly, BioPerl has modules like Bio::Seqio that enable users to handle structured data very efficiently. This is in contrast to the case of SRS and Entrez, where the user has to employ the regular expressions explicitly to retrieve tag values and other information. This makes SRS and Entrez very slow and more difficult to use in comparison to BioPerl. In Perl script using Entrez produces the protein IDs by following the expression:

*if($efetch_result=~/\s+\/protein_id=\"([^\"]+)\"/) "*

However, this is using a regular expression and assumes that the protein_id will only be once. BioPerl only requires the use of the tag name; hence the accidental incidence of the word "protein_id" in any other place does not affect the results.

*$feat_object->get_tag_values('protein_id')*

Additionally, in the case of BioPerl, the network utilization is very efficient. For multiple IDs, it is necessary to access the web interface just once to acquire all the details in a file in the required format. However, both SRS and Entrez require the initiation of a separate connection for each ID, while in Perl script a network connection is initiated for every geneid.

*my $efetch = "$utils/efetch.fcgi?". "rettype=$report&retmode=text&".*
*"db=$db&id=$id[$retstart]";*
*my $efetch_result = get($efetch); if (! defined $efetch_result)*
*{*
*die "Cannot fetch data from NCBI EntrezeFetch";*
*}*
*$efetch_result=~/LOCUS\s+\S+\s+(\d+) bp/;*

37

In the case of BioPerl, only one network requires connection.

*my $factory = Bio::DB::EUtilities->new(-eutil => 'efetch',*
*-db => 'protein', -rettype => 'gb', -id => \@ids);*
*$factory->get_Response(-file => $file);*

In SRS and Entrez, the users program was strongly coupled with the information format and hence the hard coding makes the code more prone to future failures. However, in BioPerl, the code is loosely coupled and structured data is analyzed without use of hard coding, which makes the programs more reliable. In Perl script for translation, we are used hard coded word "translation".

*$efetch_result=~/translation=\"([^\"]+)\"/;*

However, in the case of BioPerl, we can use simple tag value translation, as follows:

*get_tag_values('translation')*

According to the obtained results, the key features of structured data tools, i.e., SRs and Entrez portals presented data in the form of tables with web links for further manual exploration, whereas semi-structured tools, i.e., Perl and BioPerl programs presented data in a convenient and automatic process, which speed up the investigation technique for the user. In addition, the difference between data sets retrieved by SRS and Entrez tools, and by Perl and BioPerl can distinguish the following:

- The way, in which the queries are expressed. SRS and Entrez allow the users to start the search with keywords, whereas Perl and BioPerl program allows the user to request precise and flexible query conditions.

- The way the data found by the study is returned. In the case of SRS and Entrez, data is presented in structured format (tables, with web links for further manual exploration), whilst Perl or BioPerl programs can retrieve data in XML (semi-structured) form suitable for further processing.

## 4. RESULTS AND DISCUSSION

### 4.1 ANALYSIS OF RESULTS

The results acquired in this study prove that SRS is a powerful tool relying on keywords and verifying that it is possible to use this tool to make simple, powerful queries. The user needs to retrieve the results from the IDs query and click on each ID description to display a mix information related with this ID, such as text or numbers as required. If a user wishes to query about genes and gather specific information or calculate related statistics, the user needs to search for each access number and manually compute the statistic operation, which is time consuming.

Entrez has also been proven to be a powerful tool with simple queries, but one that has severe limitations with respect to retrieving subs-entities that need to be manually extracted. Users also need to click to each ID to display with detail the information associate with the ID. It seems fair to say that SRS and Entrez work well with simple keywords but not with keywords collected of a number of terms, and that both have problems with complex queries, which combine several search terms. Both tools have limitations with respect to retrieving subsequences, which are not automatically extracted and take a long time for users to obtain.

Both Perl and BioPerl scripts have been used to extract the same information and found that these tools are able to effectively retrieve semi-structured data. Perl gives access to data stored in GenBank database via a flexible series of sequences. BioPerl is useful to retrieve information as semi-structured data, because it is able to handle queries without knowing data types and provides flexible querying for data that allows a user to request condition in their queries to enable expected to fetch answer, which match with original query. Instead of using a web portal, the data retrieval by Perl and BioPerl was returned in XML through these programs provided by bioinformatics organizations, called EntrezeUtils tools. It was also found that when using BioPerl script, the multiple IDs called the web interface just once to acquire all the required details in a file in the required format. However, in Perl script network connection was initiated for every gene ID, in comparison to the case of SRS and Entrez, where we had to initiate a connection separately for each ID. BioPerl code was also found to require much less use of hard coding, making the programs more reliable. In Perl script for translation, we used hard coded word "translation". Perl and BioPerl allow capturing as little or as much of the full data structure as the users wish.

Comparing these two approaches to retrieve information related with these groups of proteins, we have found that the semi-structured data tools seem to have a superior performance to structured data tools, in terms of both the time required and the ability to answer the user requirement. For example, if users need to know author name for each protein ID through structured data tools, they must manually access each number to retrieve this information, which is potentially very time consuming. In contrast, using the semi-structured data tools answer the users query directly on the screen, and allow data to be copied when required. This high speed and accuracy, in addition to semi-structured data integrating the bulk of information, which does not fit easily into a set of database tables or does not conform with structure of database tables, makes the program methods clearly superior.

The analysis of results of this study shows that its limitation makes Entrez unfit for its intended purpose. SRS could be employed to index phrases, but has not been done, perhaps because utilizing an exact phrase is only appropriate, if the user already knows that phrase. In contrast, most of the time users have complex queries. Thus, the access to bioinformatics databases by Entrez, its abilities for retrieving the exact data set are limited and subsequences are not automatically extracted. SRS could be employed to index phrases, but the proprietor has not yet done this.

A script using Perl has been shown to be flexible and precise, allowing users to specify a search order in the annotations with extremely precise queries through the command line, combining time of extraction with precision. Using BioPerl as an alternative tool for gathering semi-structured data proved that, it was able to reduce the complex analyses of Perl program through typing few lines of Perl code. In addition, it deals well with specific data required by the user. Both Perl and BioPerl have been proven to be flexible enough to support the semi-structured form. Semi-structured data obtained in this way is particularly suitable for further exploration, processing and integration with other data in which the user is interested.

The challenges met when retrieving information from databases, especially when dealing with structured data that include many issues associated with the integration of complex design of many structured databases and integration approaches. This reinforces the fact that there are several technical challenges in data integration. Most biological databases utilize diverse DBMSs and do not provide a standard way of accessing the data. Some databases provide huge text dumps of their contents, whereas others permit access to the original DBMS and others still offer only web pages as their primary means of access.

### 4.2 A COMPARISON OF DATA RETRIEVAL TOOLS

According to all the evidences and examples, which mentioned in this study, we have compared the two methods to access bioinformatics databases, using structured data and semi-structured data retrieval tools, as follows:

- **Types of structures:** Structured data retrievals are usually associated with relational databases. Almost all commonly DBMSs are constructed for structural data. However, semi-structured data should be provided electronically from database systems, file systems such Web data.

- **Access:** Structured data that are public free-access databases, such as NCBI, GenBank, whereas BioPerl languages programming, such as Perl are open sources project.

- **Accuracy of data retrieved:** Structured data is useful when retrieving data from remote databases electronically from websites but not helpful when it retrieves full structures. However, semi-structured data is useful when retrieving data from remote database electronically from website and helpful as response to users inquiry immediately, and has efficient processing engines and inquiry.

- **Time:** Structured data sometimes time consuming, because it requires dealing with the full structure, whereas semi-structured data deal with specific data, which is required by the user so it may save time compared with structured data.

- **Distributed evaluation:** Structured data is useful when applied to massive or unlimited numbers of files. However, semi-structured data is useful only when applied to limited number of files.

- **Facilitate for users:** In structured data, the users can find facilities and answer their inquiry through contacting staff associated with managing the database and reading about help topics available in these databases. Yet, reading tutorials, documentation and news regarding the latest versions available from open sources projects.

The most critical recommendation produced by this study is that dealing with these tools requires users to already know certain information and instructions, such as what are the database tables, what they contain, how they can be joined and how to retrieve information from more than one table. In order to take advantage of BioPerl, a user requires a basic understanding of Perl programming language, including understanding of how to use Perl modules, references, objects and methods. This means that the users should always attempt to develop their knowledge by reading tutorials, documentation and news relating to developments from open sources projects.

### 5. CONCLUSION

This study has been conducted to assess available methods for accessing public structured data and semi-structured data stored in bioinformatics databases. Two key methods have been discussed, with examples for retrieving information, including their weaknesses and strengths in terms of evaluating performance through time required and accuracy of results. In structured data, the tools used were SRS and Entrez. In semi-structured data, the search tools were BioPerl and Perl. The study has shown that, it is clear that the process of handling genomes is better handled by the use of computer software scripts as Perl and BioPerl, which have been proven to accelerate the process for retrieval through provision of flexible ways to query data that allow users to match data precisely. An advantage of retrieving data in semi-structured form is that it is suitable for processing by the user's own software, perhaps enhanced through the use of programs such as Perl and BioPerl. In contrast, the structured data can present difficulties in an integration with other data for further analysis. Even if the semi-structured data tools retrieved happens to contain a complete set of attributes for all entities, and could therefore be converted into structured data set, the possibilities for easier integration with other data remain as an advantage of the semi-structured approach. While structured relational databases are an effective, well established method for organizing and retrieving data from single databases, users of bioinformatics data should not underestimate the difficulties of using them when attempting to integrate data from multiple sources. A better approach may be to use software tools that are able to interpret and present data made available in semi-structured formats. Some more research might be conducted in order to eliminate the failures presently afflicting management of bioinformatics data through accessing online databases through stable versions of computer languages such as Perl, Java and query languages that reflect up to date information while avoiding local copies of unwanted data.